\documentclass[11pt,a4paper]{article}
\pdfoutput=1
\usepackage{graphicx}
\usepackage{dcolumn}
\usepackage{bm}

\usepackage{amsmath}
\usepackage[english]{babel}
\usepackage{lscape}
\usepackage{multirow}
\usepackage{upgreek}
\usepackage{verbatim}

\usepackage{jinstpub}

\title{A sub-micron resolution, bunch-by-bunch beam trajectory feedback system and its application to reducing wakefield effects in single-pass beamlines}

\author[a]{D. R. Bett}
\author[a]{P. N. Burrows}
\author[a]{C. Perry}
\author[a]{R. Ramjiawan}
\affiliation[a]{John Adams Institute for Accelerator Science at University of Oxford,\\Denys Wilkinson Building, Keble Road, Oxford OX1 3RH, United Kingdom}

\author[b]{N. Terunuma}
\author[b]{K. Kubo}
\author[b]{T. Okugi}
\affiliation[b]{High Energy Accelerator Research Organization (KEK),\\1-1 Oho, Tsukuba, Japan}

\emailAdd{douglas.bett@physics.ox.ac.uk}

\abstract{A high-precision intra-bunch-train beam orbit feedback correction system has been developed and tested in the ATF2 beamline of the Accelerator Test Facility at the High Energy Accelerator Research Organization in Japan. The system uses the vertical position of the bunch measured at two beam position monitors (BPMs) to calculate a pair of kicks which are applied to the next bunch using two upstream kickers, thereby correcting both the vertical position and trajectory angle. Using trains of two electron bunches separated in time by 187.6~ns, the system was optimised so as to stabilize the beam offset at the feedback BPMs to better than 350~nm, yielding a local trajectory angle correction to within 250~nrad. The quality of the correction was verified using three downstream witness BPMs and the results were found to be in agreement with the predictions of a linear lattice model used to propagate the beam trajectory from the feedback region. This same model predicts a corrected beam jitter of c.~1~nm at the focal point of the accelerator. Measurements with a beam size monitor at this location demonstrate that reducing the trajectory jitter of the beam by a factor of 4 also reduces the increase in the measured beam size as a function of beam charge by a factor of c.~1.6.}


\arxivnumber{2008.12738}

\date{\today}

\begin{document}
\maketitle
\flushbottom

\section{\label{s:Intro}Introduction}

\begin{figure}
\centering
\includegraphics[width=0.75\columnwidth]{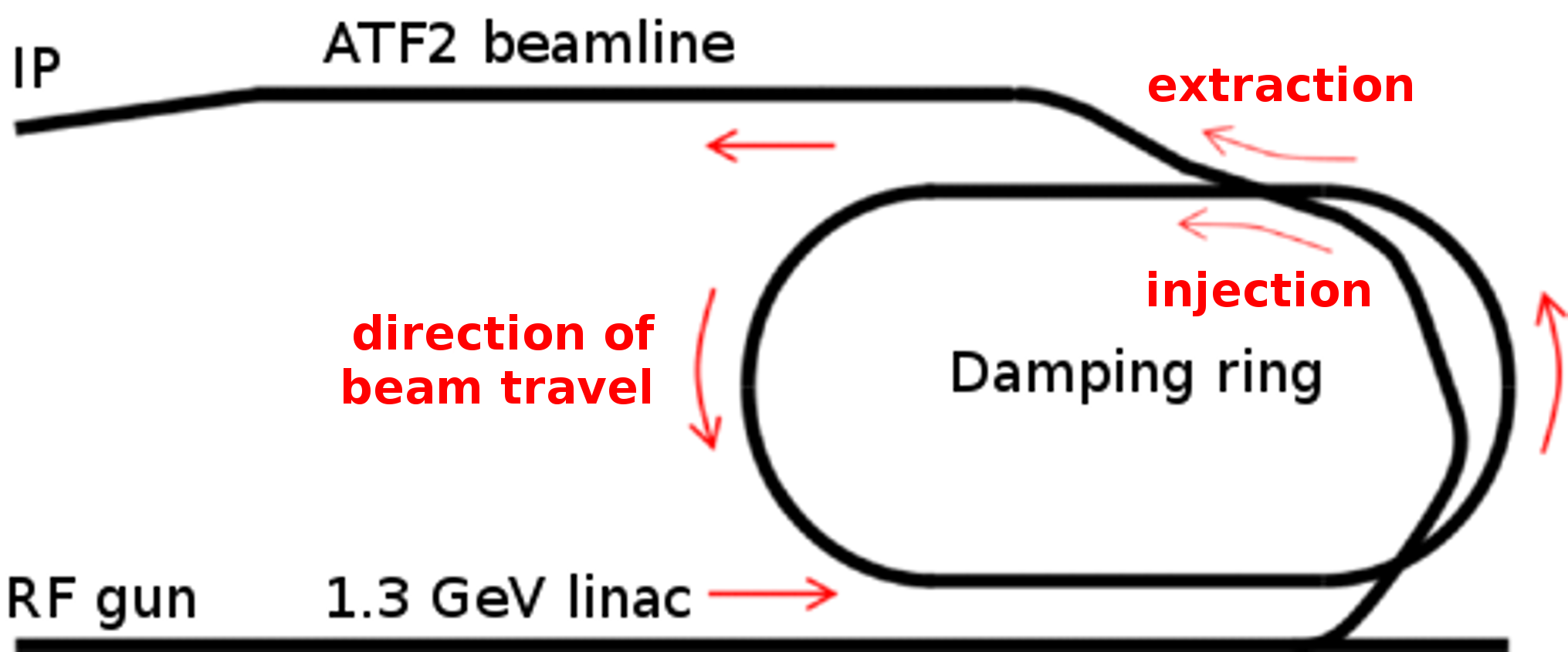}
\caption{\label{f:SchematicATF}Schematic of the ATF. The label ``IP'' refers to the nominal location of the focal point of the beamline where the beam size is minimized.}
\end{figure}

\begin{figure}
\includegraphics[width=\columnwidth]{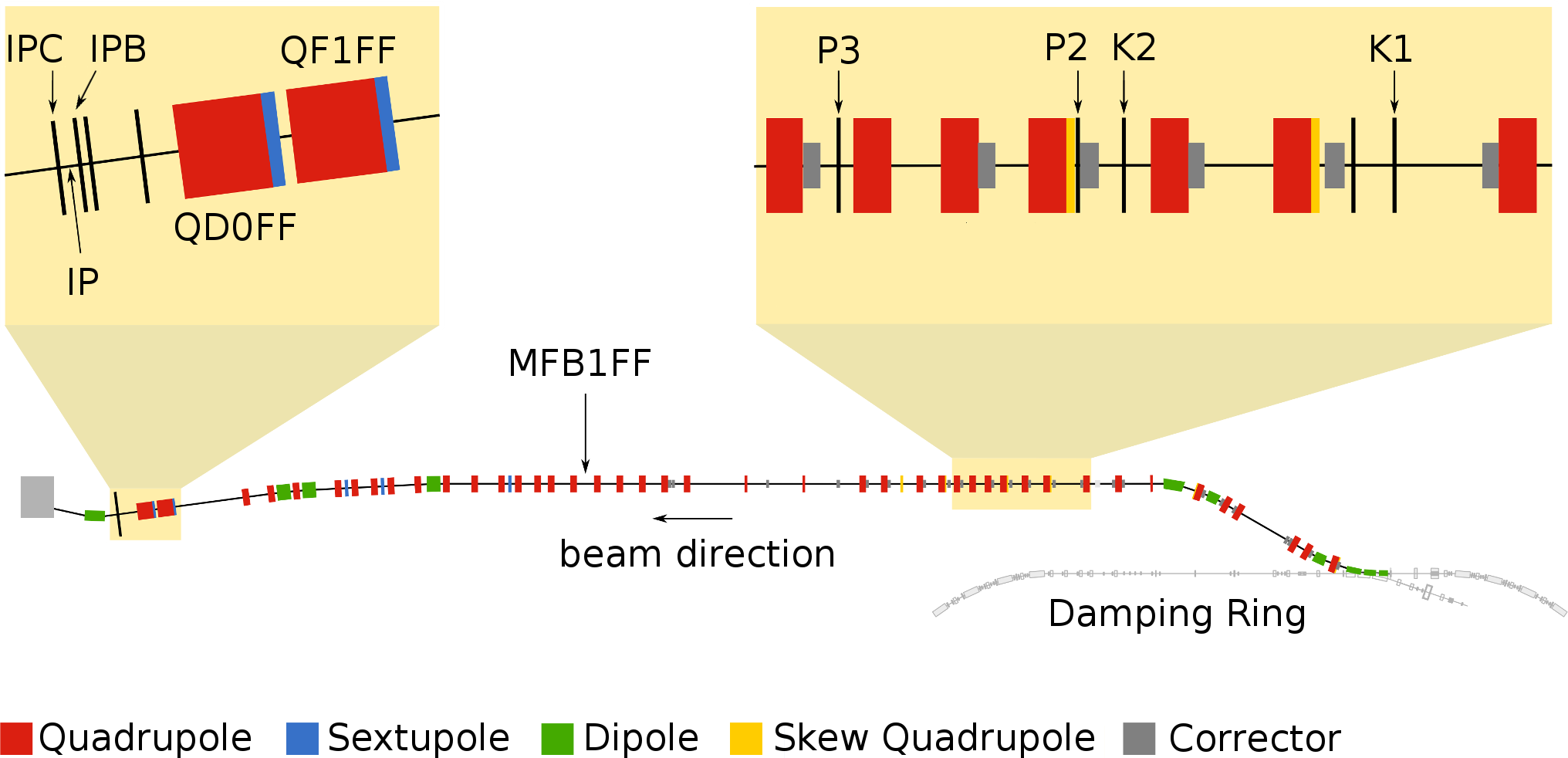}
\caption{\label{f:SchematicATF2}Schematic~\cite{Burrows:IPAC2015-MOPTY083} of the ATF2 beamline showing the layout of components in the region of the FONT feedback system and at the IP. See Table~\ref{t:BeamlineS} for location of components used.}
\end{figure}

The Accelerator Test Facility (ATF) is a research facility located at the High Energy Accelerator Research Organization (KEK) in Tsukuba, Japan. The ATF is intended to facilitate the development of technologies and techniques required for the realization of a future linear electron-positron collider, either the International Linear Collider (ILC)~\cite{JAI-2013-001} or the Compact Linear Collider (CLIC)~\cite{CERN-2018-010-M}. The ATF is shown schematically in Figure~\ref{f:SchematicATF}; it consists of an RF gun, a 1.3~GeV electron linac, a damping ring, and a beamline known as ATF2~\cite{SLAC-R-771,SLAC-R-796}. At the end of the ATF2 beamline, a pair of powerful quadrupole magnets is used to focus the electron beam to the smallest size possible at a location known as the interaction point (IP). The ATF2 beamline is shown in more detail in Figure~\ref{f:SchematicATF2}.

The ATF2 Collaboration has two goals. Goal 1 is the production of a 37~nm vertical beam spot size at the IP. Goal 2 is the stabilization of the vertical beam position at the same location to the nanometer level~\cite{Bambade:2010zz,PhysRevLett.112.034802}. The ATF is nominally operated with a beam charge of $1\times10^{10}$~electrons per bunch and a pulse repetition rate of 3.12~Hz, where each pulse consists of a single bunch with a length of approximately 7~mm~\cite{SLAC-PUB-8846}.

The ATF is also capable of generating multi-bunch trains by accumulating bunches in the damping ring over the course of several pulses and then extracting them in a single pulse. These trains consist of either two or three bunches with a bunch spacing of around 150~ns. The Feedback On Nanosecond Timescales (FONT)~\cite{URL_FONT} group at the University of Oxford developed a low-latency ($\sim$150~ns) single-phase beam feedback system~\cite{PhysRevAccelBeams.21.122802} as a prototype of the intra-train beam stabilisation system required for the interaction point of the ILC. Here, we report the results of a feedback system based on this technology to stabilize both the beam position and the trajectory angle in the ATF2. The corrections were applied in the vertical plane locally in the early part of the ATF2 beamline so as to deliver a stable beam to the entrance of the final focus system.

\begin{table}[b]
	\begin{center}
		\caption{The longitudinal locations of selected beamline components relative to the start of the ATF2 beamline.}
		\label{t:BeamlineS}
		\begin{tabular}{|r|c|}
		\hline
		Name		&	Distance [m]	\\
		\hline
		K1			&	26.672	\\
		K2			&	29.598	\\
		P2			&	30.123	\\
		P3			&	33.025	\\
		MFB1FF	&	58.534	\\
		IPB			&	89.212	\\
		IP			&	89.299	\\
		IPC			&	89.386	\\
		\hline
		\end{tabular}
	\end{center}
\end{table}

\section{\label{s:ExperimentalSetup}Feedback system}

\subsection{\label{ss:FeedbackSystem}Hardware}

\begin{figure}
\centering
\includegraphics[width=0.75\columnwidth]{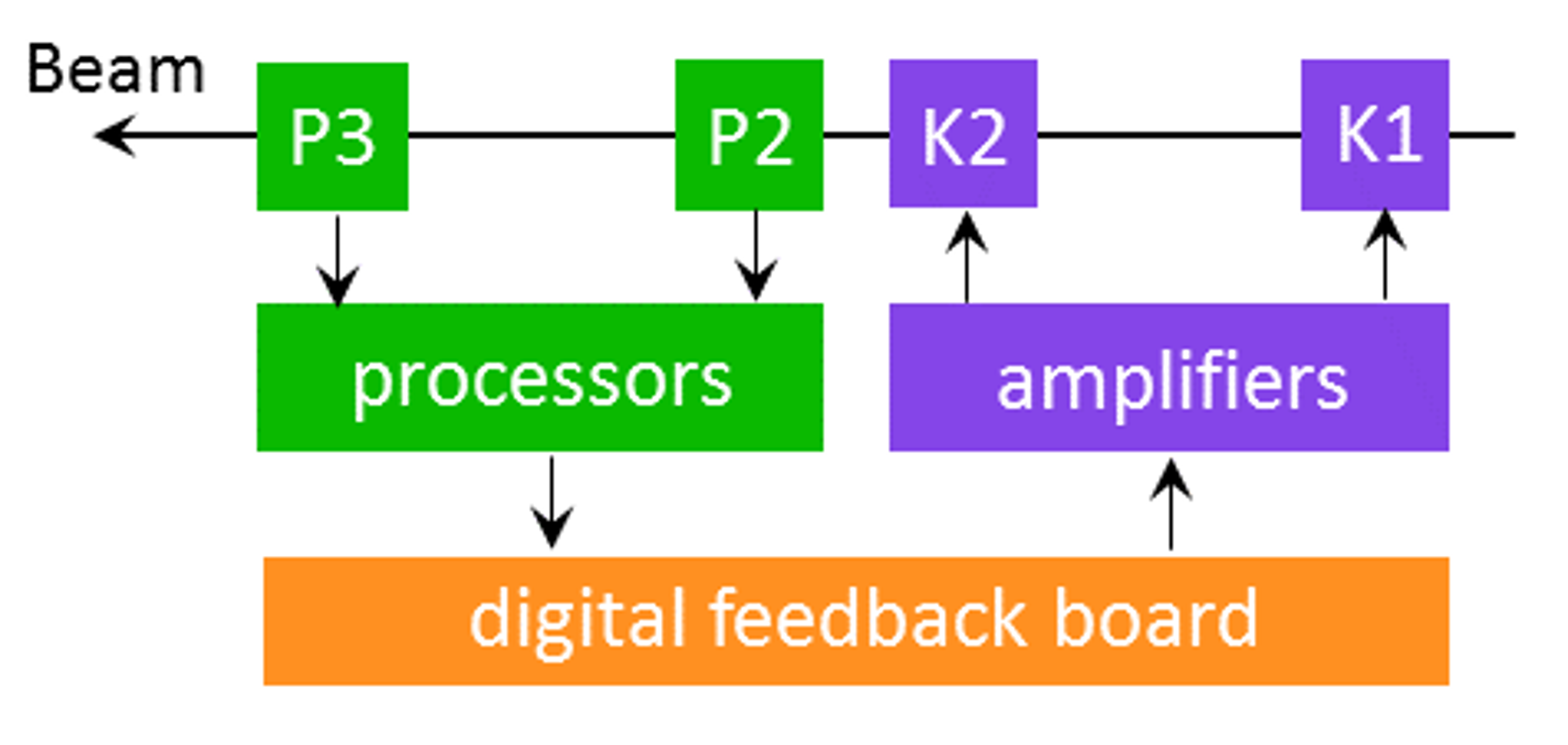}
\caption{\label{f:SchematicFont}Schematic of the coupled-loop feedback system using BPMs P2 and P3 and kickers K1 and K2.}
\end{figure}

\begin{figure}
\includegraphics[width=\columnwidth]{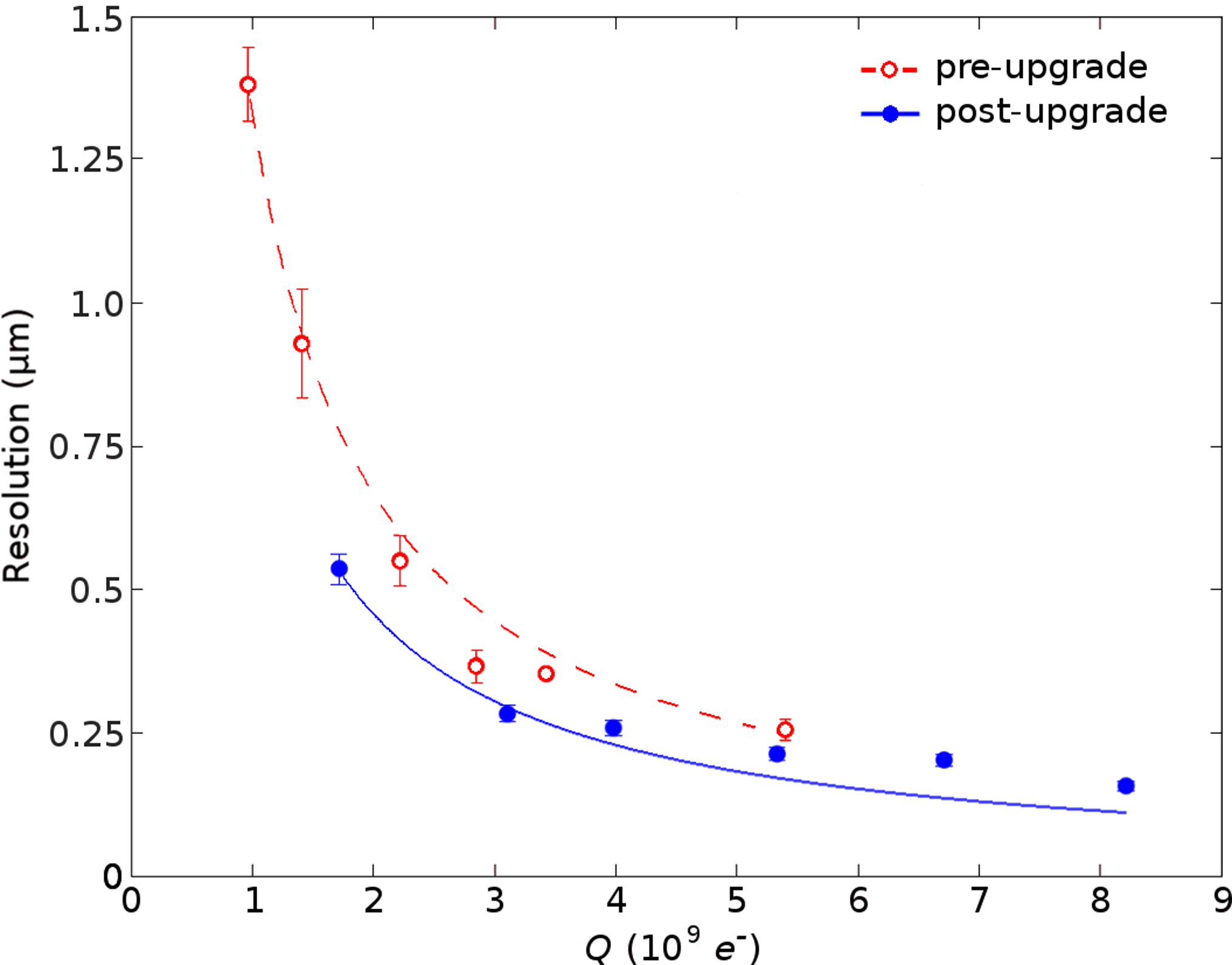}
\caption{\label{f:PlotRes}BPM resolution vs. beam bunch charge ($Q$). The filled and unfilled data points correspond to measurements with the upgraded and original systems respectively. In each case the line shows the result of extrapolating the lowest-charge data point to higher charges with a $1/Q$ dependence.}
\end{figure}

The hardware of the feedback system is depicted schematically in Figure~\ref{f:SchematicFont} and the locations of the key components relative to the start of the ATF2 beamline are given in Table~\ref{t:BeamlineS}. P2 and P3 are stripline beam position monitors (BPMs). The voltage pulses induced on the top and bottom striplines by the passage of an electron bunch are processed using custom analogue electronic modules; the design of these BPMs and electronics has been previously reported~\cite{PhysRevSTAB.18.032803}. The stripline voltage-difference signal ($\Delta$) depends on both the vertical position of the bunch and its charge $Q$, while the stripline voltage-sum signal ($\Sigma$) depends only on charge. The position of the bunch is derived from the ratio $\Delta/\Sigma$. A beam position resolution of $291\pm10$~nm for this system in operation at ATF2 has been reported~\cite{PhysRevSTAB.18.032803}. In 2016 the system was upgraded~\cite{BlaskovicKraljevic:IPAC2017-TUPIK110}, resulting in an improved position resolution of $157\pm8$~nm (Figure~\ref{f:PlotRes}) for a beam charge of 1.3~nC ($0.82\times10^{10}$~electrons/bunch). The upgraded system was used for the results reported here.

The processed BPM signals are input to a custom-made digital feedback (‘FONT5’) board~\cite{PhysRevAccelBeams.21.122802,PhysRevSTAB.18.032803}. The FONT5 board design features a Field-Programmable Gate Array (FPGA) along with nine analogue-to-digital converters and a pair of digital-to-analogue converters. The feedback algorithm runs on the FPGA and is able to calculate the appropriate kicker drive signals from the digitized BPM signals. The kicker drive signals are then amplified externally using bespoke ultra-fast amplifiers developed by TMD Technologies~\cite{URL_TMD} and applied to the stripline kickers K1 and K2. Further details of this system are reported in~\cite{Christian:PCaPAC2016-FRFMPLCO05,BlaskovicKraljevic:IBIC2016-TUPG15,BlaskovicKraljevic:IPAC2016-THPOR034,Burrows:IPAC2015-MOPTY084,Christian:IPAC2014-TUPME009}.

\subsection{\label{ss:CorrectionCalculation}Correction calculation}

For each train of two bunches extracted from the damping ring, the feedback calculation converts the measured position of the first bunch at the feedback BPMs P2 and P3 ($y_{2}$ and $y_{3}$ respectively) into a pair of kicker drive signals to be applied to the second bunch at the kickers K1 and K2 ($v_{1}$ and $v_{2}$ respectively). The derivation of the calculation is straightforward. The corrected position of the second bunch at P2, $\Phi_{2}$, is expressed as:

\begin{equation}
\Phi_{2} = Y_{2} + H_{12} v_{1} + H_{22} v_{2} \\
\label{e:Corrected}
\end{equation}

\noindent where $Y_{2}$ represents the `natural' position of bunch 2 (i.e. the position that bunch 2 would have in the absence of a kick). The second and third terms correspond to the change in position caused by the kicks at K1 and K2 respectively with $v_{i}$ representing the magnitude of the kick at K$i$ and $H_{ij}$ the kicker sensitivity constant that describes how a kick at K$i$ is converted into a position offset at P$j$. A similar expression is obtained for the corrected position of the second bunch at P3 and the two can be expressed together in a single matrix equation:

\begin{equation}
\left(
	\begin{array}{c}
	\Phi_{2} \\
	\Phi_{3}
	\end{array}
\right)
=
\left(
	\begin{array}{c}
	Y_{2} \\
	Y_{3}
	\end{array}
\right)
+
\left[
	\begin{array}{cc}
		H_{12}	&	H_{22} \\
		H_{13}	&	H_{23}
	\end{array}
\right]
\left(
	\begin{array}{c}
		v_{1} \\ v_{2}
	\end{array}
\right)
\label{e:CorrectedPosition}
\end{equation}

The goal of the feedback system is to stabilize the position of the second bunch at both BPMs i.e. $\Phi_{2} = \Phi_{3} = 0$. By imposing this condition, and assuming the upstream trajectory of bunch 2 matches that of bunch 1 ($Y_2 = y_2$), the following expression for the kicks is obtained:

\begin{equation}
\left(
	\begin{array}{c}
	v_{1} \\
	v_{2}
	\end{array}
\right)
=
-
\left[
	\begin{array}{cc}
		H_{12}	&	H_{22} \\
		H_{13}	&	H_{23}
	\end{array}
\right]^{-1}
\left(
	\begin{array}{c}
	y_{2} \\
	y_{3}
	\end{array}
\right)
\label{e:Kick}
\end{equation}

The calculation is implemented~\cite{Constance:thesis,Bett:thesis} in the firmware of the FONT5 digital board in the form:

\begin{equation}
\left(
	\begin{array}{c}
	v_{1} \\
	v_{2}
	\end{array}
\right)
=
\left[
	\begin{array}{cc}
		G_{21}	&	G_{31} \\
		G_{22}	&	G_{32}
	\end{array}
\right]
\left(
	\begin{array}{c}
	y_{2} \\
	y_{3}
	\end{array}
\right)
+
\left(
	\begin{array}{c}
	\delta v_{1} \\
	\delta v_{2}
	\end{array}
\right)
\label{e:FeedbackCoefficients}
\end{equation}

\noindent where the feedback coefficients $G_{ji}$ represent the extent to which the measured offset at P$j$ contributes to the kick to be delivered at K$i$. The feedback coefficients are derived from the measured kicker sensitivity constants $H_{ij}$, which are constant for a given set of beam optics. The $\delta v_{i}$ terms are constant offsets that can be optionally applied to the kicks, allowing the mean position of the corrected bunch to be shifted without affecting the reduction in position jitter that can be achieved.

\section{\label{s:Results}Results}

This section presents the results from two separate studies. The first study examined the beam position stability that could be achieved with the feedback system, using downstream BPMs to witness the correction. The second study explored the effect of the feedback system on the beam size at the ATF2 focal point.

\subsection{\label{ss:BeamStability}Beam stability}

The beam stability study was performed using trains of two bunches extracted from the Damping Ring with a bunch spacing of 187.6~ns, a train repetition rate of 1.56~Hz and a bunch population of $0.45 \times 10^{10}$ electrons.

\subsubsection{\label{sss:WitnessBPMs}Witness BPMs}

The stripline BPM MFB1FF (Figure~\ref{f:SchematicATF2}) is located about 25~m downstream of the feedback system (Table~\ref{t:BeamlineS}) and was instrumented with an analogue processor of the same type as used for P2 and P3. The outputs of this processor were monitored using a second FONT5 board operating purely as a digitizer. 

The cavity BPMs IPB and IPC~\cite{Jang:IPAC2016-THOAA02} (Figure~\ref{f:SchematicATF2}) are located either side of the focal point. These BPMs were instrumented with a completely distinct set of processing electronics~\cite{PhysRevSTAB.11.062801}, the outputs of which were monitored by a third FONT5 board. The operation of these BPMs for multi-bunch intra-train readout has been previously reported~\cite{Bromwich:thesis,Ramjiawan:thesis,Bromwich:IPAC2018-TUZGBD5,Ramjiawan:IPAC2018-WEPAL025}.

\subsubsection{\label{sss:BeamStabilityResults}Measurements}

\begin{figure*}
    \centering
    \begin{minipage}[t]{0.48\textwidth}
    	\centering
    	\includegraphics[width=\textwidth]{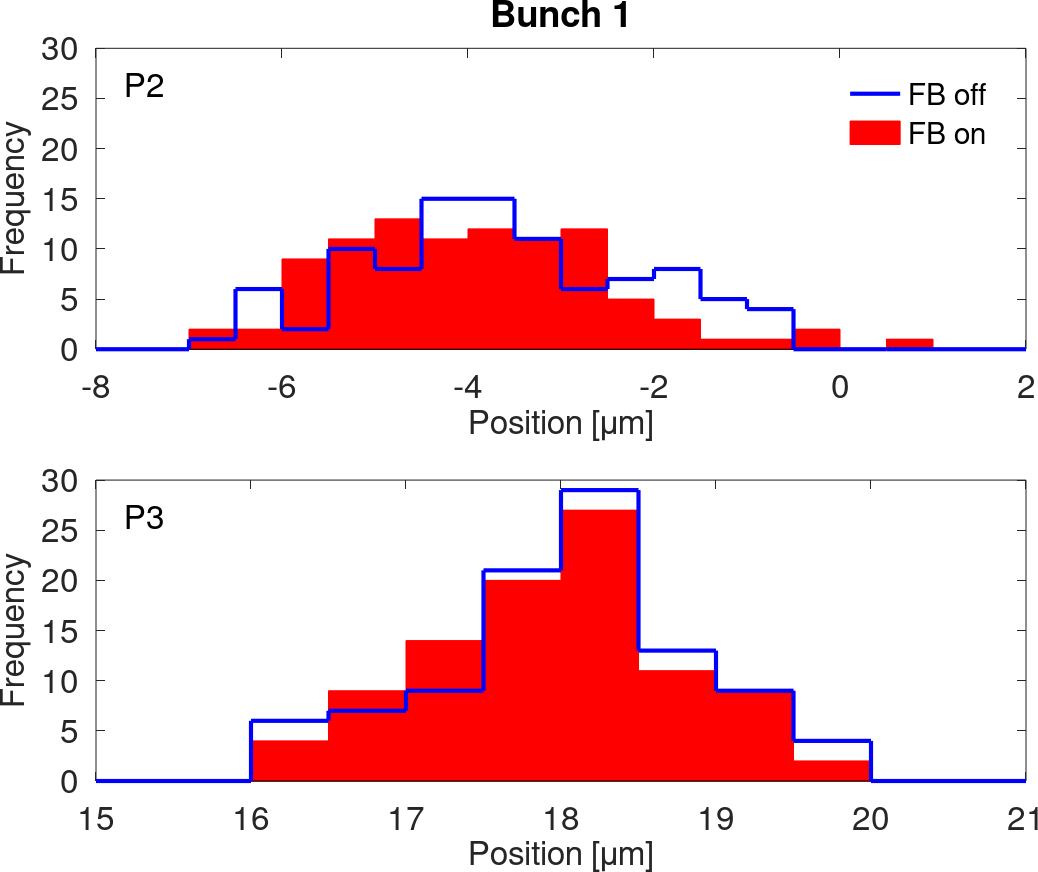}
    \end{minipage}\hfill
	\begin{minipage}[t]{0.48\textwidth}
	    \centering
    	\includegraphics[width=\textwidth]{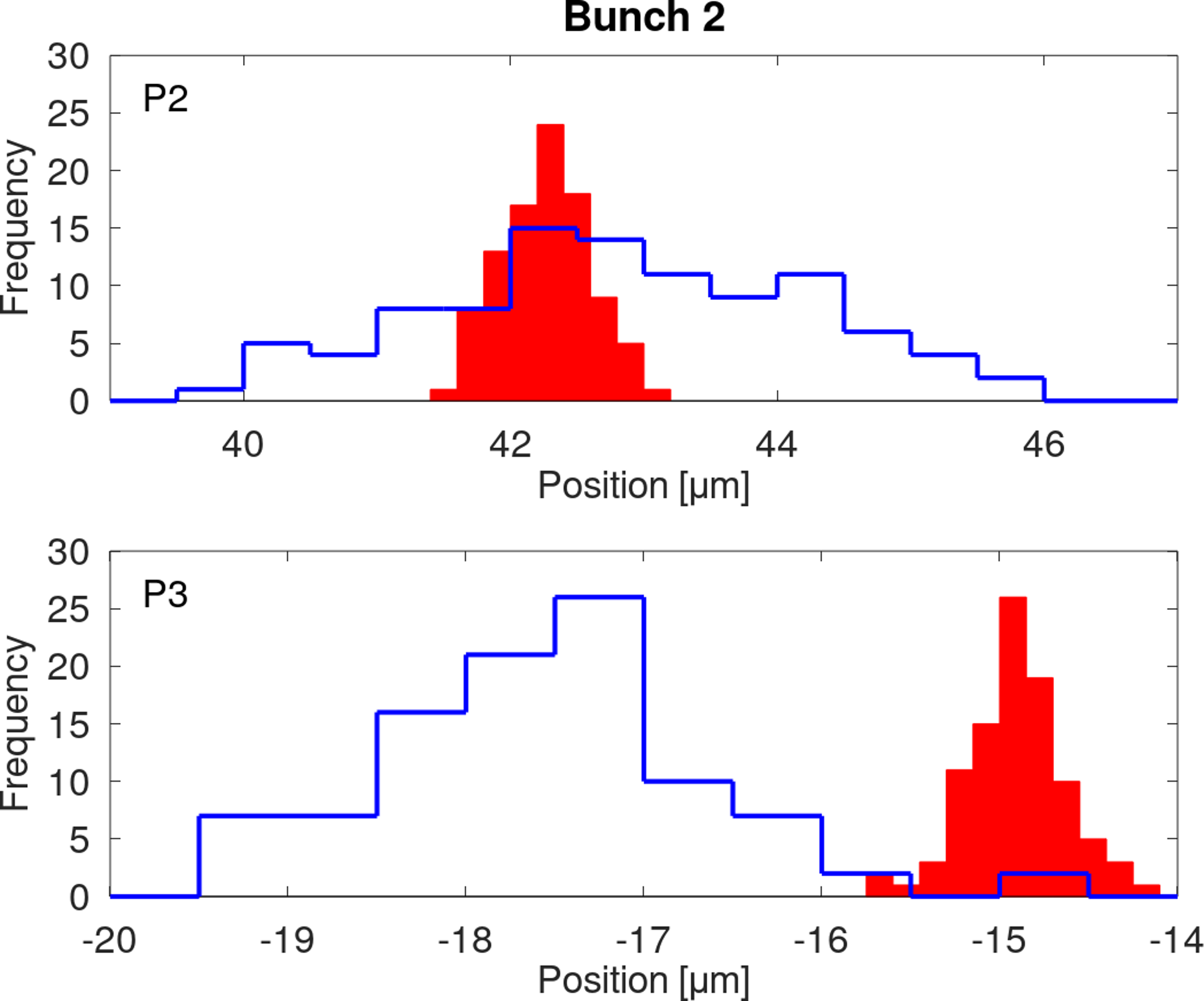}
	\end{minipage}
		
	\vspace{10pt}
	\begin{minipage}[t]{0.48\textwidth}
    	\centering
    	\includegraphics[width=\textwidth]{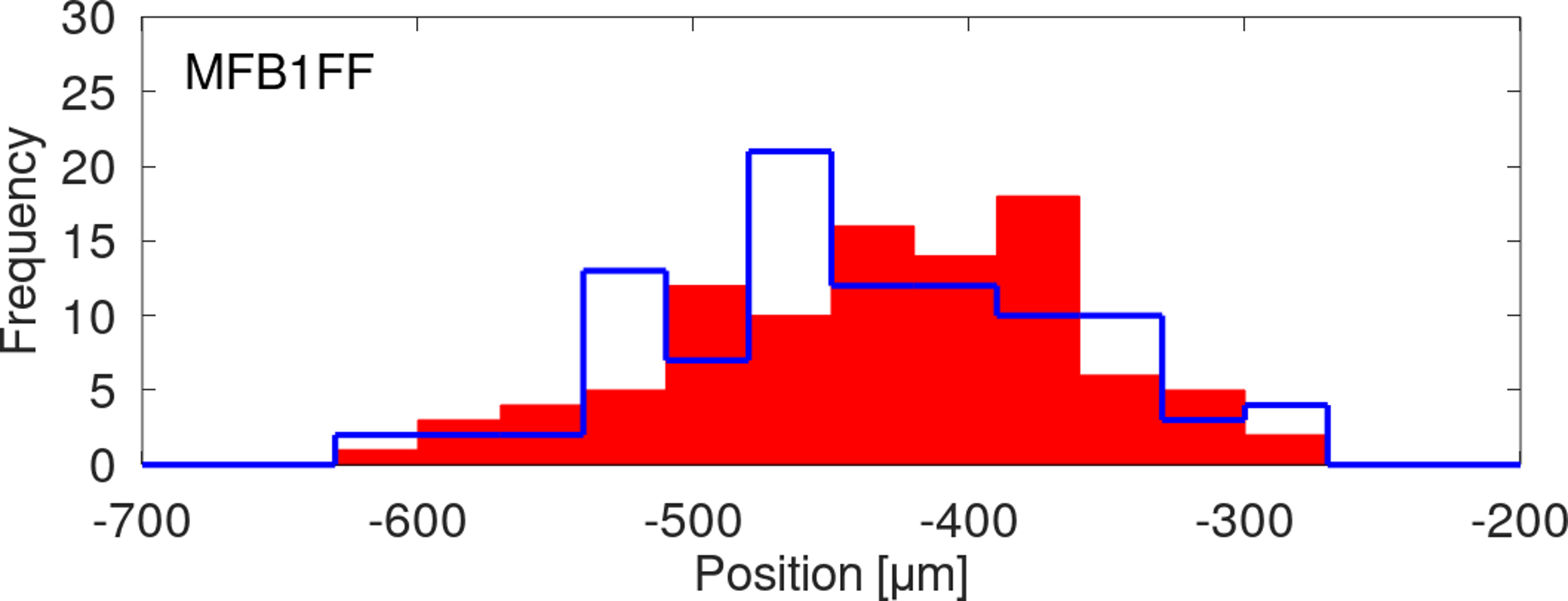}
    \end{minipage}\hfill
	\begin{minipage}[t]{0.48\textwidth}
	    \centering
    	\includegraphics[width=\textwidth]{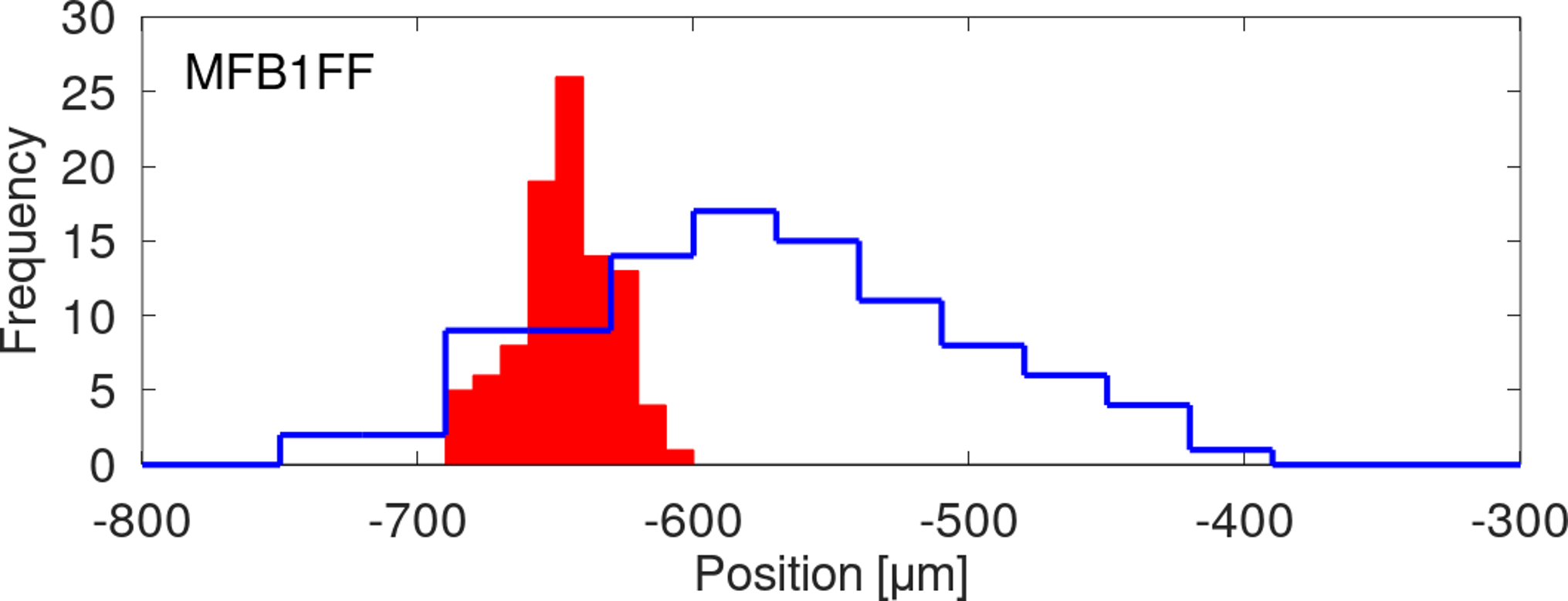}
	\end{minipage}

	\vspace{10pt}	
    \begin{minipage}[t]{0.48\textwidth}
    	\centering
    	\includegraphics[width=\textwidth]{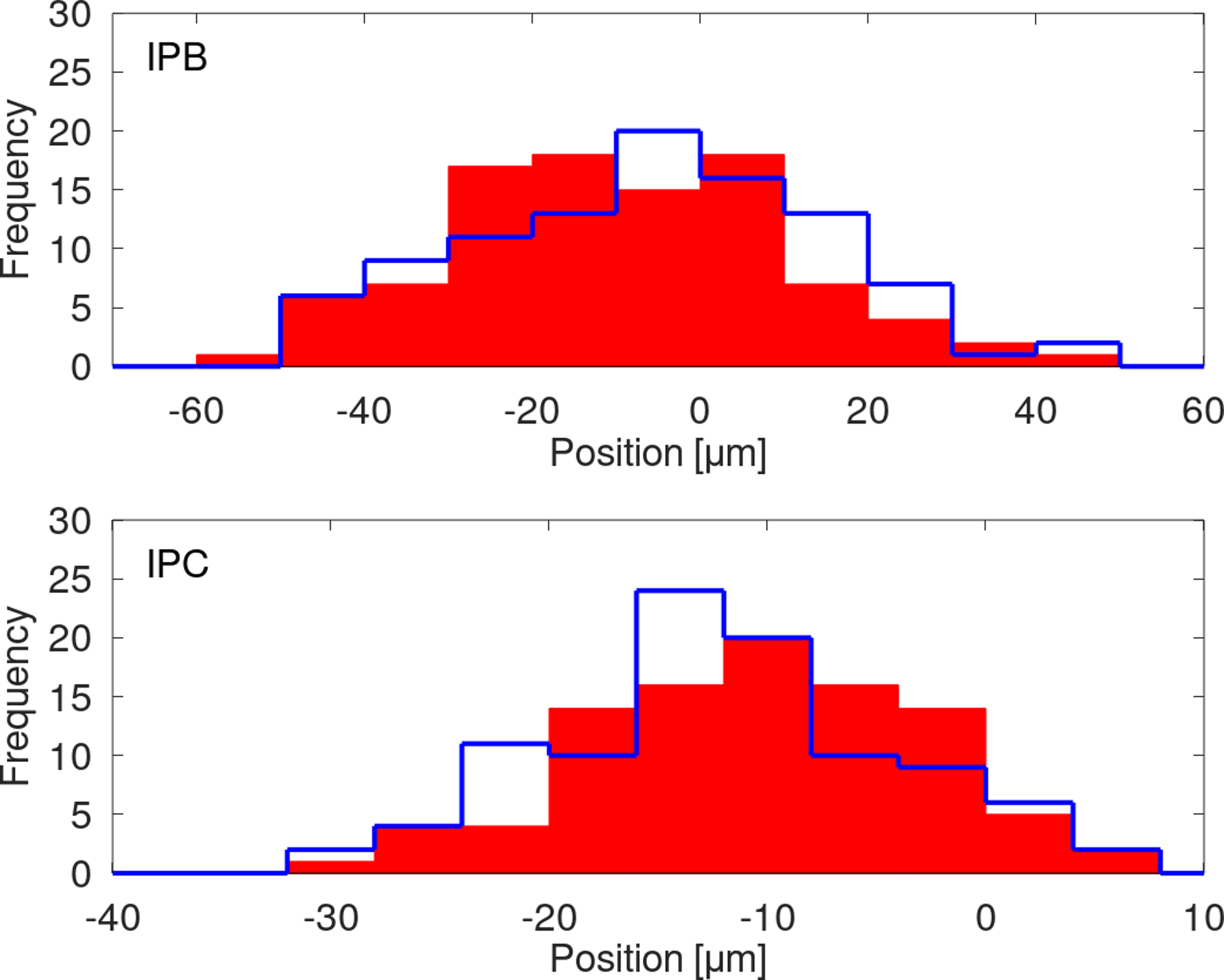}
    \end{minipage}\hfill
	\begin{minipage}[t]{0.48\textwidth}
	    \centering
    	\includegraphics[width=\textwidth]{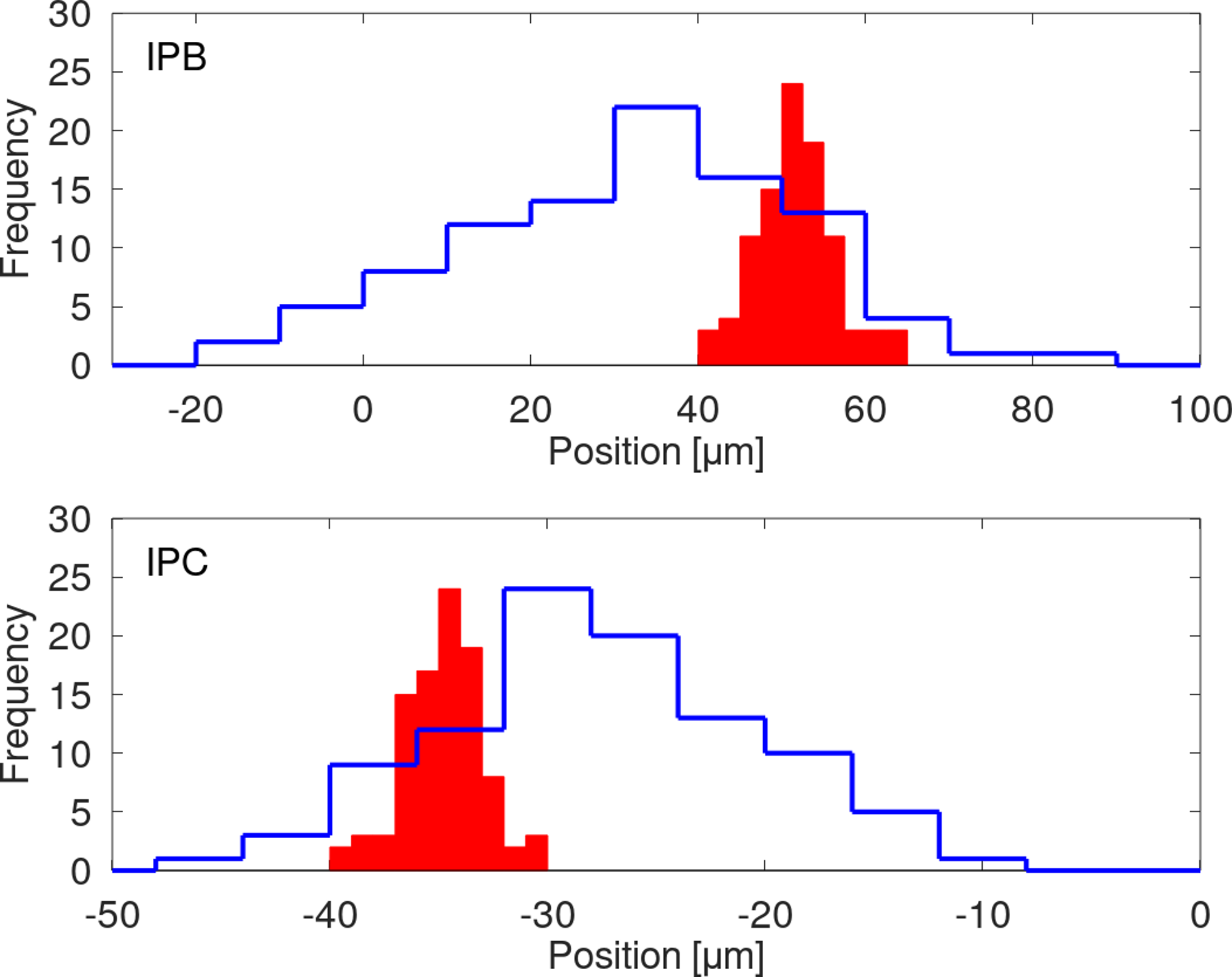}
	\end{minipage}	
	\caption{\label{f:PosHist}Distribution of position measured at each BPM (rows) for bunch 1 (left column) and bunch 2 (right column) with feedback off (outline) and on (filled). Where necessary a reduced bin width is used to display the feedback-on data so as to limit the maximum frequency of a single bin for display purposes.}
\end{figure*}

Distributions of the vertical beam position recorded at each BPM are shown in Figure~\ref{f:PosHist} for a typical run comprising 200 beam pulses. The feedback was toggled on and off for alternate beam pulses and the distributions are shown separately for the feedback-off and feedback-on sets of pulses. 

The feedback BPMs themselves are mounted on translatable mover stages and, at the start of a period of data taking, are normally aligned so as to approximately zero the mean of the readout position of bunch 1. It is clear from the feedback-off data that there is a difference of $\sim35~\upmu$m in the orbits of the two bunches, suggesting a non-uniformity of the extraction kicker pulse that removed the bunch train from the damping ring. The relative timing of the extraction kicker pulse can be adjusted to ensure that neither bunch is close to the pulse edges, but the goal of this scan is to maximize the bunch-to-bunch correlation rather than match the mean orbits. The higher the correlation between the pulse-by-pulse positions of the two bunches, the more stable the position of the corrected bunch is. The kick offset parameters (Eq.~\ref{e:FeedbackCoefficients}) are available to eliminate the residual offset of the mean position at each BPM.

In this case, the requirement to keep the corrected trajectory of the second bunch within the dynamic range of the downstream witness BPMs (including MFB1FF, which has no mover) complicates the issue and the set of measured mean positions represents the end result of an iterative process of tuning the corrected orbit of the second bunch while working within the limits imposed by the incoming orbit difference of the uncorrected bunches and the range of the cavity BPM movers.

\begin{figure*}
    \centering
    \begin{minipage}[t]{0.48\textwidth}
    	\centering
    	\includegraphics[width=\textwidth]{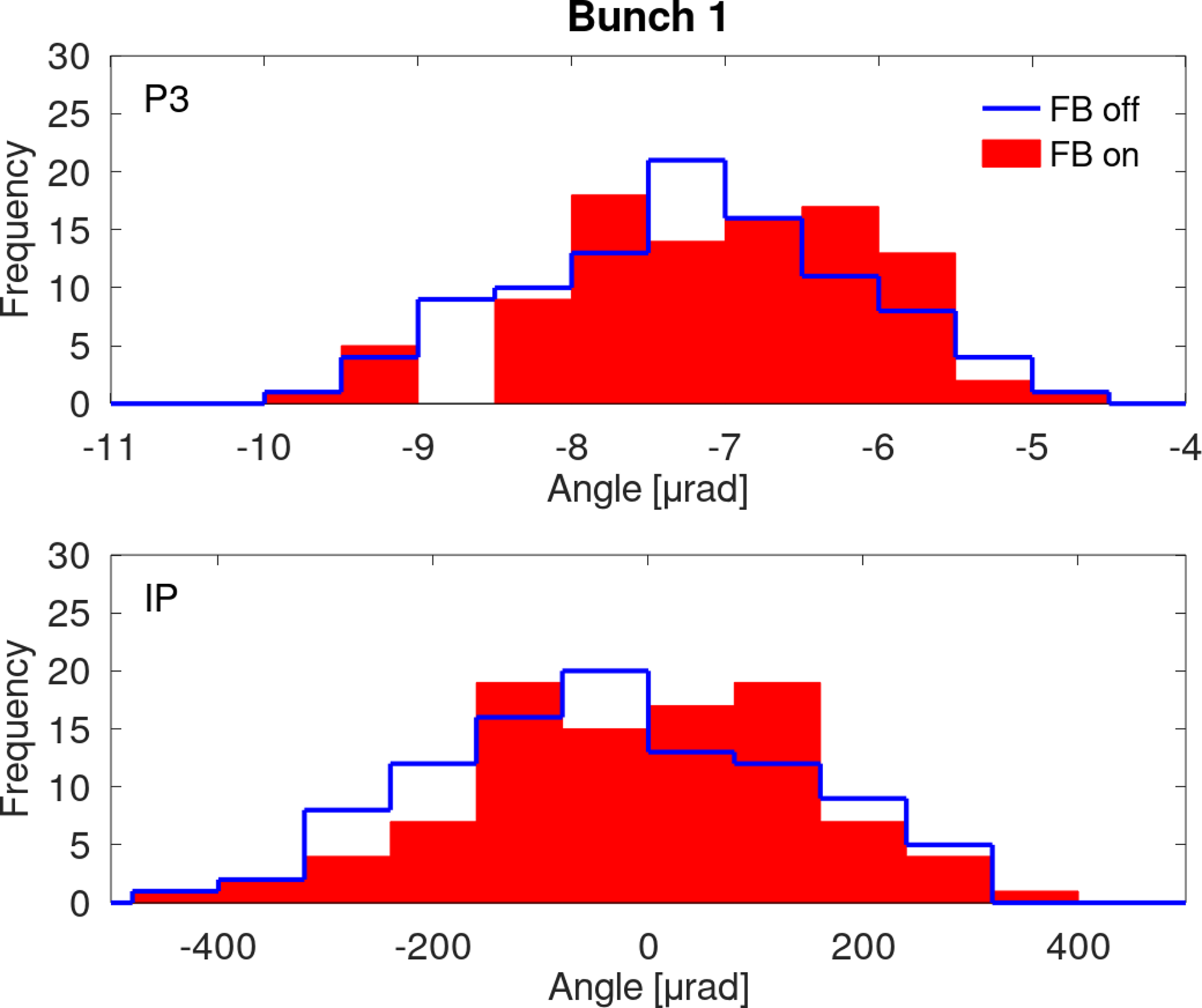}
    \end{minipage}\hfill
	\begin{minipage}[t]{0.48\textwidth}
	    \centering
    	\includegraphics[width=\textwidth]{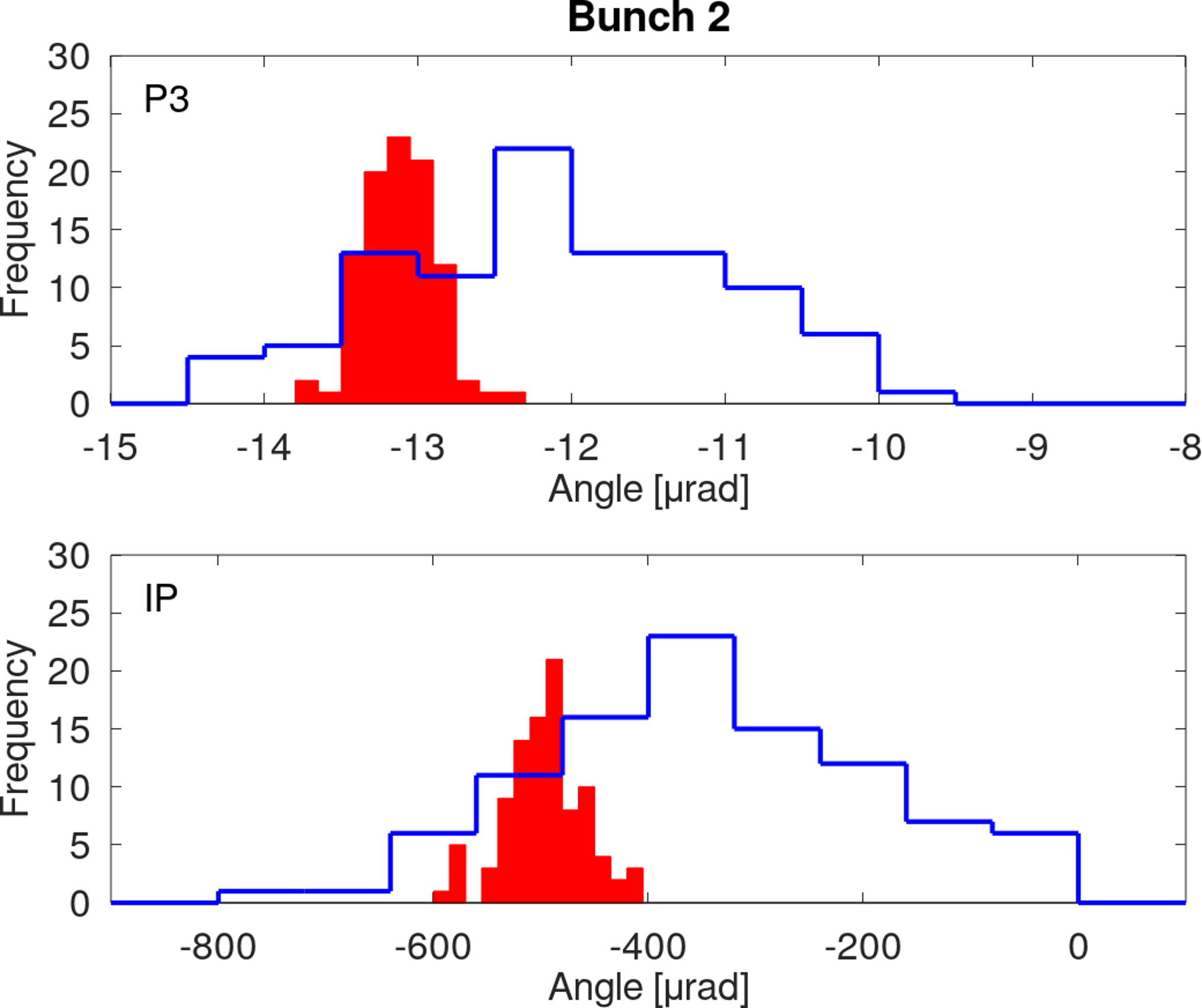}
	\end{minipage}		
	\caption{\label{f:PlotAng}Distribution of angle at P3 (calculated from the position at P2 and P3) and in the IP region (calculated from the position at IPB and IPC) with feedback off (outline) and feedback on (filled). A reduced bin width is used for the feedback on data where necessary to limit the maximum frequency of a single bin for display purposes.}
\end{figure*}

\begin{figure}
    \centering
    \includegraphics[width=0.75\columnwidth]{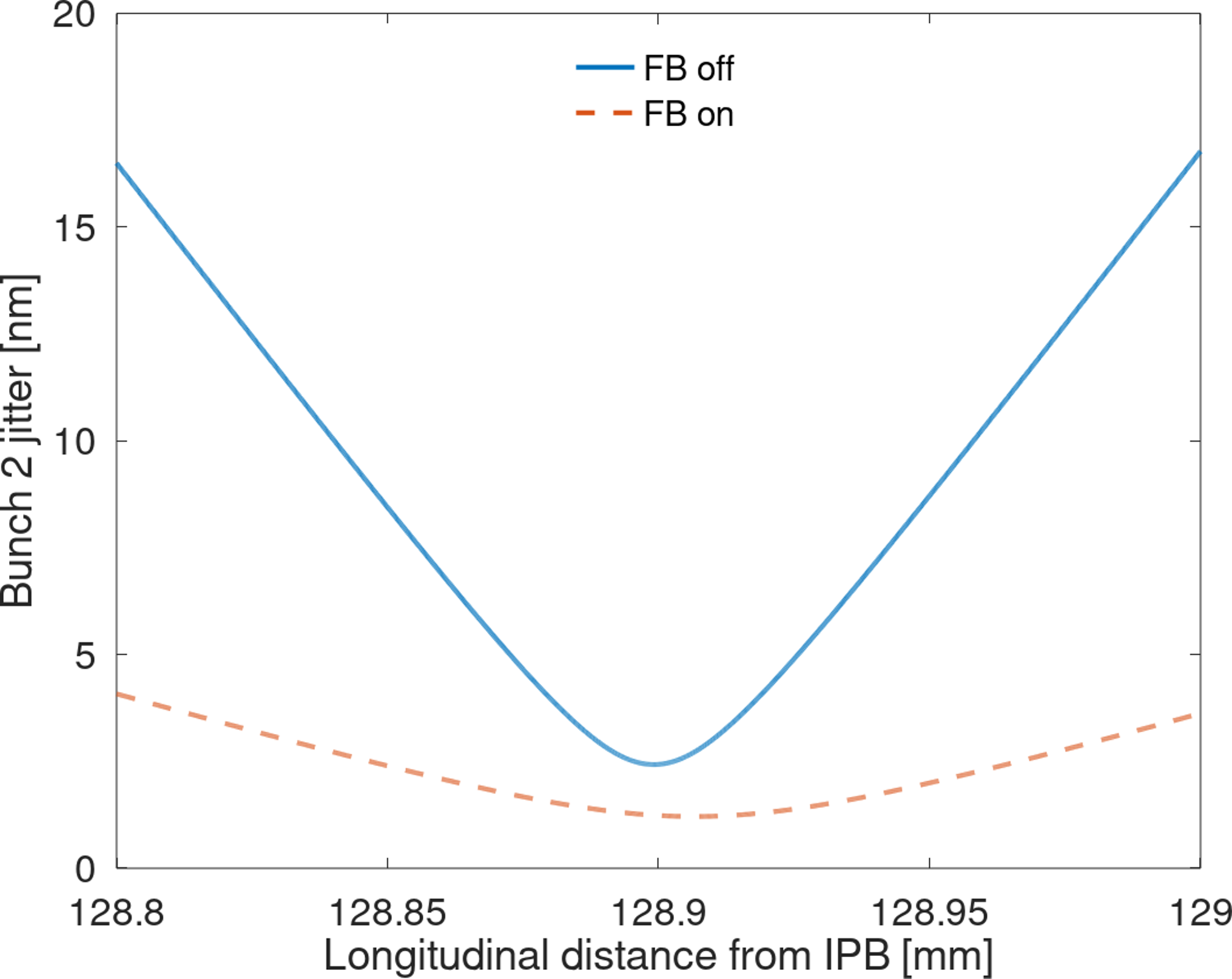}
    \caption{\label{f:PlotJitter_IP}Predicted vertical position jitter (calculated from the position at P2 and P3) in the region of the focal point with feedback off (solid) and feedback on (dashed).}
\end{figure}

\begin{figure*}
    \centering
    \begin{minipage}[t]{0.48\textwidth}
    	\centering
    	\includegraphics[width=\textwidth]{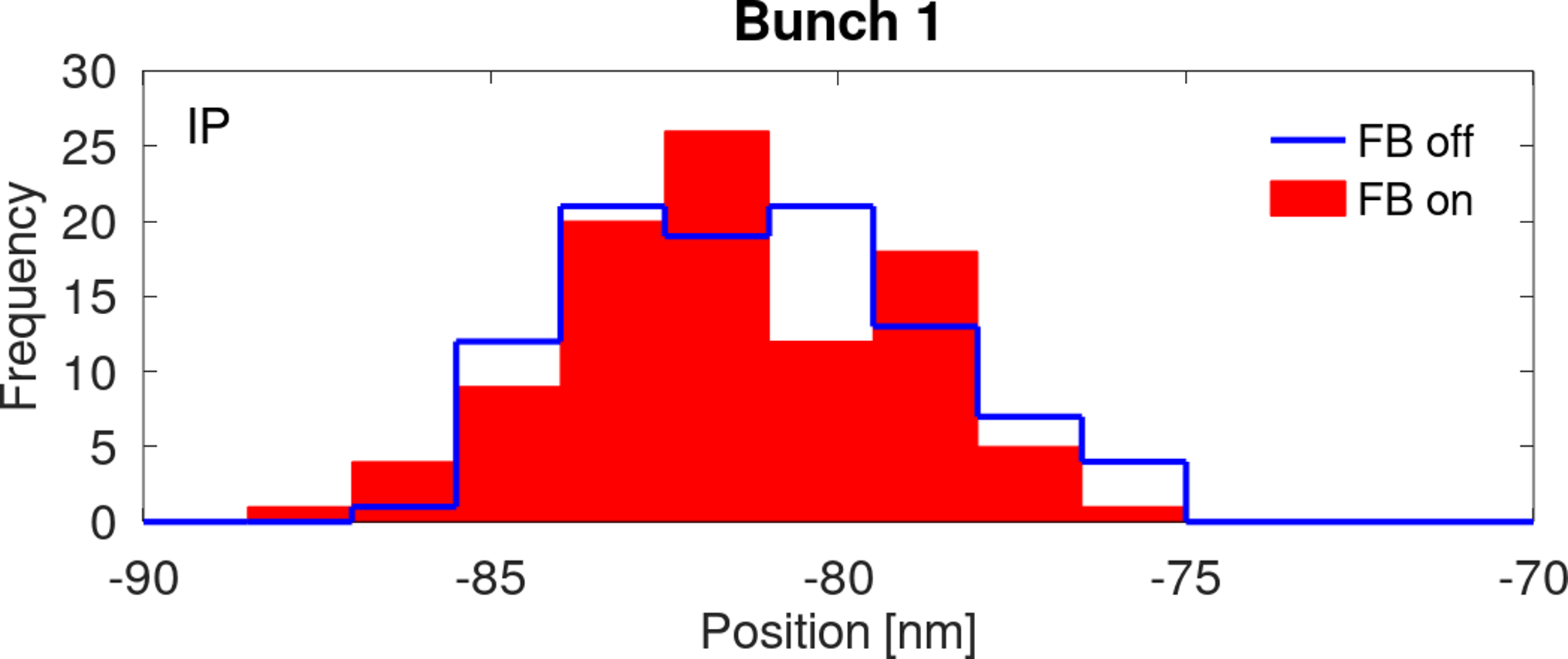}
    \end{minipage}\hfill
	\begin{minipage}[t]{0.48\textwidth}
	    \centering
    	\includegraphics[width=\textwidth]{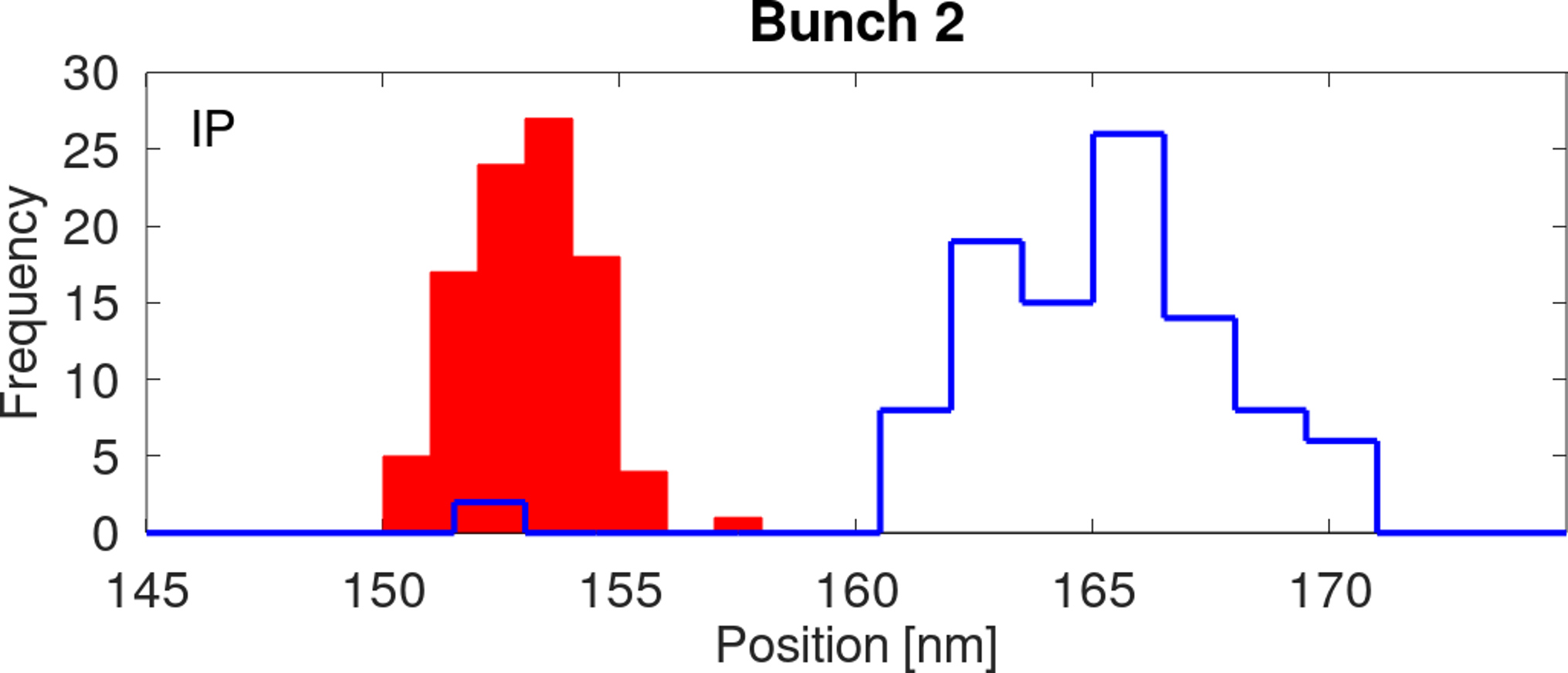}
	\end{minipage}		
	\caption{\label{f:PlotFocus}Predicted distribution of position at the focal point (calculated from the position at P2 and P3) with feedback off (outline) and feedback on (filled). A reduced bin width is used for the feedback on data where necessary to limit the maximum frequency of a single bin for display purposes.}
\end{figure*}

The performance of the feedback system in terms of the beam stability is shown in Figure~\ref{f:PosHist}. Bunch 1 provides the feedback input and its position is not corrected. Bunch 2 is well corrected by the feedback as shown by the substantial reduction in the position jitter seen at the two feedback BPMs. Table~\ref{t:Jitter} summarises the measured beam position jitter at each BPM for bunches 1 and 2 with feedback off and on, along with the correction factor, defined as the ratio of the feedback-off jitter to the feedback-on jitter. The correction is limited by the resolution of BPMs P2 and P3, which was approximately 200~nm for the bunch charge used ($0.45\times10^{10}$~electrons; see Figure~\ref{f:PlotRes}). The correction factor at all three witness BPMs is consistent with the in-loop correction of roughly a factor of 4.

Also shown in Table~\ref{t:Jitter} are the predictions of a linear beam transport model of the ATF2 beamline based on MAD~\cite{URL_MAD}. The measured beam positions at P2 and P3 were extrapolated using the model to give predicted positions at MFB1FF, IPB and IPC. The predicted jitter values and respective correction factors are in good agreement with the direct measurements, implying that there are no major sources of additional beam jitter between the feedback kickers and the ATF2 final focus.

As the system is dual-phase, the effect of the feedback on the angular jitter of the beam is also of interest. The angular jitter of the bunch is calculated using the position measured at two BPMs and knowledge of how the beam propagates from one BPM to the other; the MAD model is used for the transfer matrix from P2 to P3. In the IP region the transfer matrix is trivially obtained as the beam propagates in a ballistic fashion from IPB to IPC.

The measured position and angle can then be propagated downstream using additional transfer matrices from the model in order to give the predicted distributions of the beam angle at each witness BPM. The angles at P3 and in the IP region are shown in Figure~\ref{f:PlotAng} and these results, along with those at MFB1FF, are summarized in Table~\ref{t:AngJitter}. The results show that the angular jitter of bunch 2 is also corrected by the feedback by about a factor of 4, consistent with the position-correction analysis. The locally-measured beam angle jitter in the IP region is in good agreement with the model prediction.

The performance of the feedback system can be characterised by the degree to which it reduces the correlation between the position of bunch 1 and the position of bunch 2, and the correlation between the angle of bunch 1 and the angle of bunch 2. The calculated Pearson correlation coefficients for these two cases are shown in Table~\ref{t:Correlation} and Table~\ref{t:AngleCorrelation} respectively. The beam transport model predictions are in good agreement with the measurements. However, the data imply that the feedback is slightly over-correcting as the correlation between bunches with feedback active is slightly negative rather than consistent with zero. Optimization of the feedback coefficients to remove the residual correlation could be the subject of future studies.

The model can also be used to predict the beam position distribution at the focal point where the vertical beam position jitter is at a minimum. Figure~\ref{f:PlotJitter_IP} shows the measured jitter at P2 and P3 (Table~\ref{t:Jitter}) tracked to the focal-point region; the beam waist at the focal point is clearly visible. The tracked beam position distribution at the focal point is shown in Figure~\ref{f:PlotFocus}. With feedback off the predicted jitter is $2.9\pm0.2$~nm; with feedback operational, the equivalent jitter is $1.2\pm0.1$~nm. Figure~\ref{f:PlotJitter_IP} shows that the jitter-correction performance at the focal point is limited by the resolution of the upstream BPM inputs to the feedback.

Therefore, to the extent that the beam transport model is correct, and assuming no additional jitter sources, it is possible that the FONT feedback system corrects the beam jitter at the focal point to the level of 1~nm, thereby meeting the ATF2 beam stability goal. However, it is not possible with any known BPM technology to directly measure the beam position to the desired level of accuracy of order 1~nm, so this prediction cannot be confirmed by direct measurement. The best resolution of the cavity BPMs installed at the ATF2 IP achieved to date is c. 20~nm~\cite{Bromwich:IPAC2018-TUZGBD5}.

\begin{landscape}
\begin{table*}
	\begin{center}
		\caption{Vertical beam position jitter for both bunches for feedback off and feedback on. The top five rows are the values measured locally. The bottom three rows are the result of tracking the position data from the feedback BPMs downstream using the model. Errors are statistical.}
		\label{t:Jitter}
		\begin{tabular}{|l|l|rcl|rcl|rcl|rcl|rcl|}
		\hline
		\multirow{2}{65pt}{}											&\multirow{2}{30pt}{BPM}		&\multicolumn{6}{|c|}{Bunch 1 jitter [$\upmu$m]}						&\multicolumn{6}{|c|}{Bunch 2 jitter [$\upmu$m]}						&\multicolumn{3}{|c|}{\multirow{2}{45pt}{\centering Correction factor}} \\
																							&						 								&\multicolumn{3}{|c}{FB off}	&\multicolumn{3}{c|}{FB on}		&\multicolumn{3}{|c}{FB off}	&\multicolumn{3}{c|}{FB on}		&\multicolumn{3}{|c|}{}																									\\
		\hline
		\multirow{5}{*}{Measured locally}					&P2					 								&1.47		&$\pm$	&0.11					&1.46		&$\pm$	&0.11					&1.39		&$\pm$	&0.10					&0.34		&$\pm$	&0.02					&4.1	&$\pm$	&0.6 																											\\
																							&P3					 								&0.84		&$\pm$	&0.06					&0.81		&$\pm$	&0.06 				&0.93		&$\pm$	&0.07					&0.27		&$\pm$	&0.02 				&3.4	&$\pm$	&0.5																											\\
																							&MFB1FF			 								&74.90	&$\pm$	&5.35					&70.98	&$\pm$	&5.12					&71.89	&$\pm$	&5.13					&17.35	&$\pm$	&1.25 				&4.1	&$\pm$	&0.6																											\\
																							&IPB				 								&20.65	&$\pm$	&1.48					&19.87	&$\pm$	&1.43					&19.70 	&$\pm$	&1.41					&4.83		&$\pm$	&0.35 		 		&4.1	&$\pm$	&0.6																											\\
																							&IPC				 								&7.93		&$\pm$	&0.57					&7.57		&$\pm$	&0.55					&7.23		&$\pm$	&0.52					&1.73		&$\pm$	&0.13 		 		&4.2	&$\pm$	&0.6																											\\
		\hline
		\multirow{3}{65pt}{Tracked from P2 \& P3}	&MFB1FF			 								&77.17	&$\pm$	&5.51					&75.20	&$\pm$	&5.43					&75.54	&$\pm$	&5.40					&16.70	&$\pm$	&1.21 		 		&4.5	&$\pm$	&0.7 																											\\
																							&IPB				 								&21.77	&$\pm$	&1.56					&21.21	&$\pm$	&1.53					&21.31	&$\pm$	&1.52					&4.71		&$\pm$	&0.34 		 		&4.5	&$\pm$	&0.7																											\\
																							&IPC				 								&7.65		&$\pm$	&0.55					&7.46		&$\pm$	&0.54					&7.49		&$\pm$	&0.53					&1.66		&$\pm$	&0.12					&4.5	&$\pm$	&0.7																											\\	
		\hline
		\end{tabular}
	\end{center}
\end{table*}		

\begin{table*}
	\begin{center}
		\caption{Vertical beam angle jitter for both bunches for feedback off and feedback on. The top four rows are the result of tracking the position data from the feedback BPMs downstream using the model. The final row is obtained using the IPB and IPC position data. Errors are statistical.}
		\label{t:AngJitter}
		\begin{tabular}{|l|l|rcl|rcl|rcl|rcl|rcl|}
		\hline
		\multirow{2}{65pt}{}											&\multirow{2}{30pt}{BPM}		&\multicolumn{6}{|c|}{Bunch 1 jitter [$\upmu$rad]} 													&\multicolumn{6}{|c|}{Bunch 2 jitter [$\upmu$rad]}													&\multicolumn{3}{|c|}{\multirow{2}{45pt}{\centering Correction factor}}	\\
																							&														&\multicolumn{3}{|c}{FB off}	&\multicolumn{3}{c|}{FB on}										&\multicolumn{3}{|c}{FB off}	&\multicolumn{3}{c|}{FB on}										&\multicolumn{3}{|c|}{ }																								\\
		\hline
		\multirow{4}{65pt}{Tracked from P2 \& P3}	&P2													&1.11		&$\pm$	&0.08					&1.08		&$\pm$	&0.08 												&1.08		&$\pm$	&0.08 				&0.24		&$\pm$	&0.02 												&4.6	&$\pm$	&0.7																											\\
																							&P3													&1.07		&$\pm$	&0.08					&1.04		&$\pm$	&0.08 												&1.04		&$\pm$	&0.07					&0.23		&$\pm$	&0.02 												&4.5	&$\pm$	&0.7																											\\			
																							&MFB1FF											&29.91	&$\pm$	&2.14					&29.15	&$\pm$	&2.10													&29.29	&$\pm$	&2.09					&6.48		&$\pm$	&0.47													&4.5	&$\pm$	&0.7																											\\	
																							&IP													&168.89	&$\pm$	&12.06				&164.58	&$\pm$	&11.88 												&165.34	&$\pm$	&11.81				&36.56	&$\pm$	&2.64  												&4.5	&$\pm$	&0.7																											\\
		\hline
		Measured locally													&IP													&164.05	&$\pm$	&11.72				&157.49	&$\pm$	&11.37												&154.54	&$\pm$	&11.04				&37.38	&$\pm$	&2.70													&4.1	&$\pm$	&0.6																											\\
		\hline
		\end{tabular}
	\end{center}
\end{table*}
\end{landscape}

\begin{table}
	\begin{center}
		\caption{Bunch-to-bunch position correlation coefficient for feedback off and feedback on. The top five rows are the values measured locally. The bottom three rows are the result of tracking the position data from the feedback BPMs downstream using the model. Errors are statistical.}
		\label{t:Correlation}
		\begin{tabular}{|l|l|rcl|rcl|}
		\hline
																									&BPM			&\multicolumn{3}{c|}{FB off}		&\multicolumn{3}{c|}{FB on} \\
		\hline
		\multirow{5}{*}{Measured locally}							&P2				&0.95	&$\pm$	&0.03							&-0.23	&$\pm$	&0.10 			\\
																									&P3				&0.90	&$\pm$	&0.04							&-0.31	&$\pm$	&0.10 			\\
																									&MFB1FF		&1.00	&$\pm$	&0.01							&-0.17	&$\pm$	&0.10 			\\
																									&IPB			&0.99	&$\pm$	&0.01							&-0.18	&$\pm$	&0.10 			\\
																									&IPC			&0.99	&$\pm$	&0.01							&-0.18	&$\pm$	&0.10 			\\
		\hline
		\multirow{3}{60pt}{Tracked from P2 \& P3}			&MFB1FF		&0.97	&$\pm$	&0.02							&-0.25	&$\pm$	&0.10 			\\
																									&IPB			&0.97	&$\pm$	&0.02							&-0.25	&$\pm$	&0.10 			\\	
																									&IPC			&0.97	&$\pm$	&0.02							&-0.25	&$\pm$	&0.10 			\\	
		\hline																							
		\end{tabular}
	\end{center}
\end{table}	

\begin{table}
	\begin{center}
		\caption{Bunch-to-bunch angle correlation coefficient for feedback off and feedback on. The top four rows are the result of tracking the position data from the feedback BPMs downstream using the model. The final row is obtained using the IPB and IPC position data. Errors are statistical.}
		\label{t:AngleCorrelation}
		\begin{tabular}{|l|l|rcl|rcl|}
		\hline
																									&BPM			&\multicolumn{3}{c|}{FB off}		&\multicolumn{3}{c|}{FB on} \\
		\hline
		\multirow{4}{60pt}{Tracked from P2 \& P3}			&P2				&0.98	&$\pm$	&0.02							&-0.25	&$\pm$	&0.10 			\\
																									&P3				&0.98	&$\pm$	&0.02							&-0.25	&$\pm$	&0.10 			\\
																									&MFB1FF		&0.97	&$\pm$	&0.02							&-0.25	&$\pm$	&0.10 			\\
																									&IP				&0.97	&$\pm$	&0.02							&-0.25	&$\pm$	&0.10 			\\
		\hline
		Measured locally															&IP				&0.99	&$\pm$	&0.01	 						&-0.18	&$\pm$	&0.10 			\\
		\hline
		\end{tabular}
	\end{center}
\end{table}

\subsection{\label{ss:BeamSize}Beam size}

In addition to its application as a direct means of achieving the beam stability goal at ATF2, the FONT beam orbit feedback system has also been observed to cause a reduction in the apparent beam size at the IP~\cite{Okugi:PASJ2016-MOOL04}. This is thought to be a result of a better-controlled beam experiencing smaller wakefield-induced distortions of the bunch shape within particular structures along the beamline.  We report the results of a study to investigate the effect of beam orbit control on the measured beam size as a function of the bunch charge.

\subsubsection{\label{sss:BeamSizeMonitor}Beam size monitor}

A nanometer-resolution IP beam size monitor (IPBSM) is installed at the ATF2 IP~\cite{Suehara01}. The device works by splitting a laser beam in two and then crossing the two halves at the IP to form a fringe pattern in the beam focal plane. The size of the fringes is given by $d=\lambda/2\sin(\theta/2)$, where $\lambda$ is the laser wavelength and $\theta$ is the crossing angle of the two laser paths. Laser photons are inverse Compton scattered by the electron beam and measured downstream of the IP. The position of the fringes relative to the beam is scanned by phase shifting one of the laser beams and the degree of variation of the scattered photon signal is quantified as a modulation depth ($M$). The vertical beam size ($\sigma$) can then be estimated from:

\begin{equation}
  \sigma = \frac{1}{k}
  \sqrt{\frac{1}{2} \ln \left( \frac{C \left| \cos \theta \right|}{M} \right)} 
\label{e:IPBSM}
\end{equation}

\noindent where $k = \pi/d$ and $C$ expresses the contrast reduction of the laser fringe pattern due to factors such as deteriorated spatial coherency of the laser.

\subsubsection{\label{sss:WakefieldEffects}Beam size growth due to wakefields}

The interaction of the electromagnetic field surrounding a bunch of charged particles with geometrical discontinuities in the beamline results in wakefields. Each particle in the bunch receives a transverse deflection from the wakefield induced in the beam pipe by the passage of the preceding particles, leading to both a change in the measured orbit of the bunch as a whole as the center of mass shifts and a change in the orbit of the tail of the bunch relative to the head. As the IPBSM effectively measures the size of the distribution of particles at the IP integrated over many bunches, any increase in the beam position jitter or distortion of the transverse profile of the bunch is perceived as an increase in beam size.

ATF2 is known~\cite{Korysko:thesis} to be particularly sensitive to wakefields due to the long bunch length and the relatively low beam energy. The primary sources of wakefields in the ATF2 beamline are C-band cavity BPMs, bellows and vacuum flanges~\cite{Okugi:PASJ2019-FRPI023}. The orbit change caused by wakefields at ATF2 has been reported~\cite{PhysRevAccelBeams.19.091002} and several of the cavity BPMs were removed in order to reduce it. As the magnitude of the wakefield kick is proportional to the position offset between bunch and wakefield source (for small offsets), a position feedback that reduced the offset between bunch and wakefield source would be expected to mitigate the increase in beam size due to wakefields. This is described in the next section.

\subsubsection{\label{sss:BeamSizeMeasurements}Measurements}

\begin{figure*}[htb]
    \centering
    \includegraphics[width=\textwidth]{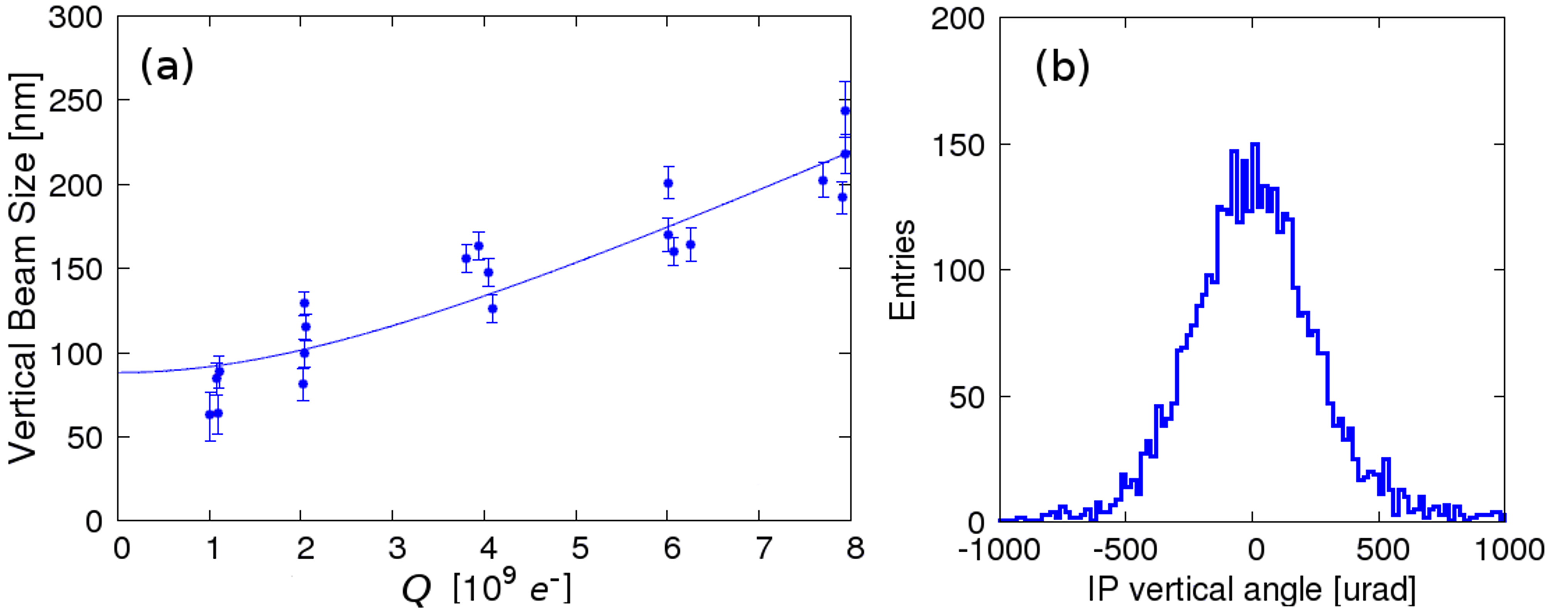}
	\caption{\label{f:PlotBeamSizeSingle}Beam size as a function of beam charge (left) and distribution of the IP vertical angle jitter (right) for single bunch operation. Each point represents a single beam size measurement. The line is a fit of Eq.~\ref{e:BeamSize}.}
	\vspace{10pt}
	\includegraphics[width=\textwidth]{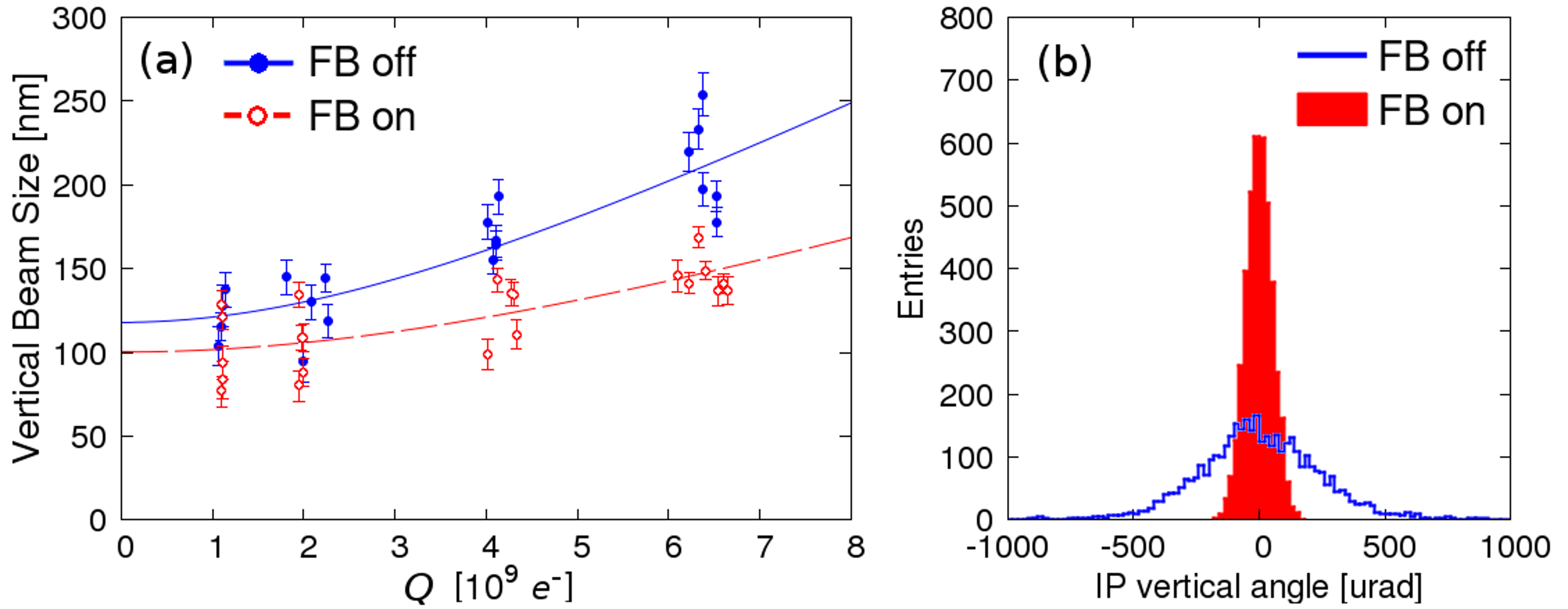}
	\caption{\label{f:PlotBeamSizeMulti}Beam size as a function of beam charge (left) and distribution of IP vertical angle jitter (right) for two bunch operation with feedback on (unfilled points, dashed line) and feedback off (filled points, solid line). Each point represents a single beam size measurement. The lines are fits of Eq.~\ref{e:BeamSize}.}
\end{figure*}

Figure~\ref{f:PlotBeamSizeSingle}a shows the beam size as a function of the beam charge when the beam was operated in single bunch mode. The vertical beam size ($\sigma$) can be expressed as a function of a charge dependence parameter ($w$):

\begin{equation}
	\sigma = \sqrt{\sigma_0^2 + w^2 Q^2}
	\label{e:BeamSize}
\end{equation}

\noindent where $\sigma_0$ is the beam size in the absence of wakefields. Fitting Eq.~\ref{e:BeamSize} to the data yields $w = 25.1\pm1.5$~nm/$10^{9}e^{-}$ (Figure~\ref{f:PlotBeamSizeSingle}a). Using measurements from the cavity BPMs in the ATF2 beamline~\cite{PhysRevSTAB.15.042801}, the IP vertical angle jitter was estimated to be 220~$\upmu$rad (Figure~\ref{f:PlotBeamSizeSingle}b) for a bunch charge between $4~\times~10^{9}$ and $6~\times~10^{9}$ electrons. 

After the measurement in single bunch mode, the ATF was set up to provide trains consisting of two bunches separated by 302.4~ns. Using the ATF2 cavity BPMs, the uncorrected vertical angle jitter of the second bunch at the IP was estimated to be $\sim215~\upmu$rad (Figure~\ref{f:PlotBeamSizeMulti}b). With the upstream feedback system active, the jitter is reduced to 51~$\upmu$rad. Figure~\ref{f:PlotBeamSizeMulti}a shows the measured size of the second bunch as a function of beam charge, both with and without feedback. It can be seen that stabilizing the position and angle of the second bunch with the FONT feedback system also reduced the charge dependence of the beam size measured at the IP by a factor of $1.6\pm0.2$, from $27.4\pm1.9$~nm/$10^9e^{-}$ to $16.9\pm1.6$~nm/$10^9e^{-}$. Table~\ref{t:BeamSize} shows a summary of the IP vertical angle jitters and the fitted charge dependence parameters of the second bunch. The magnitude of this reduction is in line with what would be expected from a detailed model of beam transport in the ATF2 beamline including explicitly the known wakefield sources; full details are reported in ~\cite{Korysko:thesis, ATFwakefieldpaper}. 

\begin{table*}
\begin{center}
\caption{Summary of charge dependence of beam size. Errors are statistical.}
\begin{tabular}{|l|c|c|}
\hline
																	&	IP angle jitter [$\upmu$rad]	&$w$ [nm/$10^{9}e^{-}$]	\\ 
\hline
Single bunch operation       			&	$220\pm16$										&$25.1\pm1.5$						\\ 
Two bunch operation without FB 		&	$215\pm15$										&$27.4\pm1.9$						\\ 
Two bunch operation with FB  	  	&	$51\pm4$											&$16.9\pm1.6$						\\ 
\hline
\end{tabular}
\label{t:BeamSize}
\end{center}
\end{table*}

\section{\label{s:Conclusions}Conclusions}

An intra-train position and angle feedback system has been developed to achieve the ATF2 beam stability goal. Operating on a train of two bunches separated by 187.6~ns, the feedback system stabilized the position of the second bunch at the feedback BPMs to the 270-340~nm level and the angle to within 250~nrad. The model of the beamline predicts that this level of correction should deliver a factor 4.5 reduction in both position and angle at the downstream witness BPMs, and the actual observed factor is $4.1\pm0.6$. The model also predicts that the jitter at the focal point should be reduced from $3.0\pm0.2$~nm to $1.2\pm0.1$~nm, meeting the target performance.

The potential of the feedback system towards reaching the beam size goal by reducing the impact of wakefields on the beam size was also measured. Scanning the charge of a beam consisting of a single bunch indicated that the beam size is increased in quadrature by $25.1\pm1.5$~nm for each additional $10^{9}$ electrons in the bunch. Repeating the charge scan using trains of two bunches separated by 302.4~ns showed that the feedback system reduced the charge dependence of the quadrature growth in beam size from $27.4\pm1.9$~nm/$10^{9}e^{-}$ to $16.9\pm1.6$~nm/$10^{9}e^{-}$.

The position and angle feedback system presented here is based on earlier development of the ILC IP collision feedback system prototype that was reported in~\cite{PhysRevAccelBeams.21.122802}. The current system was developed for specific application to beam jitter reduction at the ATF2, and, as we have shown here, as such it is successful in yielding a significant reduction in the impact of wakefields on the beam-size growth. Such a system would also be directly applicable to deployment in the ILC linacs and/or beam-delivery system for removal of correlated position/angle jitter among the c. 1300 bunches per train of each beam pulse.

\section{\label{s:Acknowledgments}Acknowledgments}

We thank KEK for their support of ATF/ATF2 operations, and KEK staff for their outstanding logistical support. In addition, we thank our colleagues from the ATF2 Collaboration for their help and support. In particular, we thank the IFIC group from the University of Valencia for providing the mover system for the feedback BPMs, the LAL group from the Paris-Saclay University for providing the mover system for the cavity BPMs at the IP, and the team from Kyungpook National University for providing the interaction point cavity BPMs.

We acknowledge financial support for this research from the United Kingdom Science and Technology Facilities Council via the John Adams Institute, University of Oxford, and CERN, CLIC-UK Collaboration, Contract No. KE1869/DG/CLIC. The research leading to these results has received funding from the European Commission under the Horizon 2020/Marie Sk\l{}odowska-Curie Research and Innovation Staff Exchange (RISE) project E-JADE, Grant Agreement No. 645479.

\clearpage

\end{document}